\documentclass[twoside]{article}
\usepackage{qic,epsfig}
\usepackage{amsmath}

\begin{document}
\setlength{\textheight}{8.0truein}    

\runninghead{Title  $\ldots$}
            {Author(s) $\ldots$}

\normalsize\textlineskip
\thispagestyle{empty}
\setcounter{page}{1}

\copyrightheading{0}{0}{2013}{000--000}

\vspace*{0.88truein}

\alphfootnote

\fpage{1}

\centerline{\bf
HIGH PERFORMANCE ERROR CORRECTION}
\vspace*{0.035truein}
\centerline{\bf FOR QUANTUM KEY DISTRIBUTION USING POLAR CODES}
\vspace*{0.37truein}
\centerline{\footnotesize
PAUL JOUGUET\footnote{paul.jouguet@sequrenet.fr}}
\vspace*{0.015truein}
\centerline{\footnotesize\it Institut Telecom / Telecom ParisTech, CNRS LTCI, 46 rue Barrault}
\baselineskip=10pt
\centerline{\footnotesize\it Paris, 75634,
France}
\vspace*{10pt}
\centerline{\footnotesize 
SEBASTIEN KUNZ-JACQUES}
\vspace*{0.015truein}
\centerline{\footnotesize\it SeQureNet, 23 avenue d'Italie}
\baselineskip=10pt
\centerline{\footnotesize\it Paris, 75013, France}
\vspace*{0.225truein}
\publisher{(received date)}{(revised date)}

\vspace*{0.21truein}

\abstracts{
We study the use of polar codes for both discrete and continuous variables Quantum Key Distribution (QKD). Although very large blocks must be used to obtain the efficiency required by quantum key distribution, and especially continuous variables quantum key distribution, their implementation on generic x86 Central Processing Units (CPUs) is practical. Thanks to recursive decoding, they exhibit excellent decoding speed, much higher than large, irregular Low Density Parity Check (LDPC) codes implemented on similar hardware, and competitive with  implementations of the same codes on high-end Graphic Processing Units (GPUs).
}{}{}

\vspace*{10pt}

\vspace*{3pt}
\communicate{to be filled by the Editorial}

\vspace*{1pt}\textlineskip    
\section{Introduction}        
With Quantum Random Number Generators (QRNGs), QKD \cite{SCA09} is among the first industrial applications of quantum information technology. The two parties of a QKD protocol, Alice and Bob, exchange quantum signals through a physical (also called quantum) channel (such as light propagation through optical fibers or free-space propagation) and can extract a secret key, secure in the information-theoretic sense, even in the presence of an eavesdropper with unlimited computational power.

Two families of QKD technologies exist: Discrete Variables (DV) QKD and Continuous Variables (CV) QKD. In both cases, the transmission of a binary message, the \emph{raw key}, on a quantum noisy channel is at the heart of the protocol. Errors resulting from this transmission have to be corrected for Alice and Bob to be able to compute the same key. The quantum channels of DVQKD and CVQKD have different error distributions: in the DVQKD case, the channel is a Binary Symmetric Channel (BSC) whose probability of error is the \emph{Quantum Bit Error Rate} (QBER). For CVQKD, it is a Gaussian channel with both a transmission $T$, and a Gaussian noise, composed of a quantum noise, the \emph{shot noise}, and other classical noises which form the \emph{excess noise}.

When linear, non-interactive, error-correcting codes are used, the error correction algorithm uses the fact that the string sent satisfies some predefined set of linear equations where some linear combinations of message bits, or \emph{parity bits}, are equal to zero. Transmission is therefore preceded by an \emph{encoding step} where the message to be transmitted is transformed into a string that satisfies these equations. However, in the QKD setting, contrary to the usual setup of error-correcting codes, a noiseless classical channel is available alongside the quantum noisy channel. Using this channel, the  encoding step can be avoided: the message and the string sent are equal, and the values of the parity bits are revealed on the classical channel. Therefore the performance of the encoding step is not considered in our case.

The limitations the error correction step introduces in the implementation of a QKD system are two-fold.
First, the number of raw key bits or linear combination of raw key bits revealed during the error correction step must be subtracted from the final key size during the \emph{privacy amplification} step \cite{Ren05thesis}. Therefore efficient codes, i.e. codes with thresholds close to the Shannon Bound, are needed. Secondly, the throughput of the error-correction, which is usually not high because of the aforementioned efficiency constraints, may limit the final key rate below what is allowed by the optics. On the other hand, cost, power consumption, and latency constraints are much less of an issue than in typical error-correction applications.

We propose to examine the relevance for QKD of a new family of codes, \emph{polar codes}, introduced by Ar{\i}kan \cite{Ari08}. Based on our previous discussion, we will look at their distance to Shannon bounds and the decoding speed.

For a given block size $N$ and a fixed channel, the polar decoding algorithm is deterministic. Its execution time provably scales in $O(N\log{N})$; it also has a simple recursive structure which gives good  practical performance.
However, we will see that very large blocks are required to achieve the high efficiencies needed for QKD on the BSC or the Binary Input Additive White Gaussian Noise Channel (BIAWGNC).

The paper is organised as follows: in section \ref{sec:error_correction_qkd} the impact of the imperfection of the error-correction procedure in both DVQKD and CVQKD is detailed and the previous work is reviewed. In section \ref{sec:polar_codes_qkd} the usage of polar codes to correct errors in a QKD setup is laid out. Finally the performances of polar codes and LDPC codes are compared in section \ref{sec:exp_results}.

\section{Effect of an imperfect error correction step in QKD}
\label{sec:error_correction_qkd}
\subsection{Secret key rate and error correction}
\subsubsection{Key rate and distance of error correction to Shannon bounds}
\noindent
In a classical DVQKD setup, Alice encodes a classical bit onto the phase or the polarization of a photon and sends this photon to Bob who measures it with a Single Photon Detector (SPD) and gets a bit value. As regards CVQKD, Alice encodes continuous information onto the quadratures of the electromagnetic field and sends weak light pulses to Bob who performs either a homodyne measurement on one single quadrature or a heterodyne measurement on both quadratures. In both cases, Bob ends up with a bit string, like in a DVQKD setup, because of the finite precision of its measurement apparatus. Since this step is repeated many times, Alice and Bob are given two bit strings $x$ and $y$ after the quantum exchange.

The eavesdropper, Eve, has a quantum state $E$, generally correlated to $x$ and $y$. If we assume Alice is chosen as the reference for the establishment of a secret key, the maximal secret information shared by Alice and Bob is given by $S(x|E)$, which is the Von Neumann entropy of the variable $x$ conditionnally to Eve's knowledge (which is in general quantum). In order to compute an information-theoretic secret key rate, all the information corresponding to the errors between $x$ and $y$, $H(x|y)$ that is the conditional Shannon entropy of $x$ given $y$, is assumed to be known by Eve and is subtracted from the final key. Thus the theoretical secret key rate reads:

\begin{equation}
\label{eq:keyrate_th}
K_{th}=S(x|E)-H(x|y)
\end{equation}

This expression can be rewritten in terms of mutual informations as:

\begin{equation}
\label{eq:keyrate_th_mutual}
K_{th}=I(x:y)-S(x:E)
\end{equation}

According to the information theory, one can never extract the exact amount of mutual information $I(x:y)$ between Alice and Bob with a finite error-correcting code. That is why one introduces a factor $\beta$ which represents the reconciliation efficiency and ranges from $0$ when no information is extracted to $1$ in the theoretical perfect reconciliation scheme:

\begin{equation}
\label{eq:keyrate_real}
K_{real}=\beta I(x:y)-S(x:E)
\end{equation}

Thus an imperfect reconciliation scheme results in a reduction of the secret key rate and a limitation of the range of the protocol. With all known protocols $I(x,y) - S(x:E)$ decreases faster with the distance than $I(x,y)$ and $S(x:E)$ individually, so that the effect of $\beta < 1$ is most severe at large distances. This last effect limited the range of CVQKD protocols for a long time before specific error correcting codes were proposed \cite{LEV08, JKL11}. 

\subsubsection{Key rate and error correction computation time}
\noindent
Long-range QKD therefore needs error-correcting codes and decoding schemes enabling operation as close to the Shannon limit $\beta=1$ as possible. However, decoding close to the Shannon limit can be a computationnally demanding task; the computation time may then limit the throughput of a QKD experiment.
In \cite{LOD07}, the raw optical repetition rate is 500 kHz and the raw data rate reduces to 350 kHz because some pulses are used for synchronisation purposes and parameters estimation. Since the best reconciliation algorithm available in \cite{LOD07} is limited to about 63 000 symbols per second, only 18\% of the available symbols can be used to extract secret keys. More generally, the key rate of a practical system is affected by a factor $\alpha=D_{ECCout}/D_{ECCin}$ where $D_{ECCout}$ stands for the error-correction output rate (63 kb/s in our example) and $D_{ECCin}$ stands for the data output rate of the system used as an input for the error-correction (350 kb/s in our example).

\begin{equation}
\label{eq:keyrate_system}
K_{sys}=\alpha\left(\beta I(x:y)-S(x:E)\right)
\end{equation}

\subsubsection{Key rate and error correction frame error rate}
\noindent
The frame error rate (FER), or the probability for a message to be incorrectly decoded, is usually one of the most regarded characteristics of an error-correcting code, since failure to decode a message is usually associated with data loss in conventional data transmission scenarios, at best causing retransmission delays. However, in the quantum key distribution setting, raw key blocks incorrectly decoded are simply discarded by both the sender and the receiver. As a result, the raw key rate and final key rate are affected by a factor $(1-\operatorname{FER})$. 
Frame error rates that are unacceptable in conventional error correction applications are therefore sufficient in the QKD case. Besides, accepting a high FER enables faster error correction. Our target figure in the rest of this article is a FER of 0.1.

\bigskip
Taking into account all the previously discussed imperfections of ECC in the QKD case, the final key rate is 

\begin{equation}
\label{eq:keyrate_imp_ecc}
K=\alpha(1-\operatorname{FER})\left(\beta I(x:y)-S(x:E)\right)
\end{equation}

\subsection{Previous work}
\noindent
Most of the error-correction algorithms designed especially for DVQKD, such as Cascade \cite{BS94,CRE95,SUG00}, Winnow \cite{BLT03} or Liu's algorithm \cite{LTD03} suffer latency problems because they are highly interactive. Although the latest ones exhibit less interactivity than Cascade, it remains the algorithm most used in DVQKD experiments because it exhibits an efficiency higher than $96\%$ \cite{ELK09} over the range $[0;0.11]$ for the error probability of a standard Binary Symmetric Channel (BSC), which is the admissible range for the QBER to distribute a secret in DVQKD. The maximum reported Cascade speed is about $5.5$Mb/s with 4 threads on a quad-core processor \cite{SAS11}.

The use of Low Density Parity Check (LDPC) codes has been proposed (see for example \cite{WMU08}) and such codes have been developped for DVQKD experiments achieving efficiencies similar with Cascade over the range $[0;0.02]$ while they present a significant improvement for bit error rates above $0.02$\cite{ELK09}. As regards interactivity, LDPC codes require only one exchange contrary to Cascade which is highly interactive. Since LDPC codes are optimized for a given probability error, puncturing and shortening techniques \cite{EMM11b} can be used to extend their efficiency to a wider range and protocols allowing to reconcile information while maintaining a low interactivity have been proposed \cite{ELK10, EMM10a, EMM10b}. However, high-efficiency LDPC error-correction speed has not been investigated a lot except for CVQKD where the authors of \cite{LOD07} report a $40$kb/s speed on CPU and a $60$kb/s speed on GPU.

Modern coding techniques have mainly been used for continuous variables with Turbo-codes or LDPC codes. The main difficulty as regards continuous variables is that the best protocols known require a Gaussian modulation while the noise added by the channel is Gaussian too. Thus, one has to deal with an Additive White Gaussian Noise Channel (AWGNC) and high-efficiency error-correction is particularly hard at low Signal to Noise Ratios (SNRs) which correspond to a long operating distance for CVQKD. However, in \cite{LEV08}, the authors proposed a technique allowing to encode the information in binary variables which allows us to deal with a Binary Input (BI) AWGNC instead of the usual AWGNC. Low-rate high-efficiency multi-edge LDPC codes can be designed for this channel \cite{JKL11, RIC02} which results in a considerably extended achievable distance for CVQKD with a Gaussian modulation.

\section{Polar codes for QKD: efficiency vs. block sizes}
\label{sec:polar_codes_qkd}
\noindent
The use of polar codes has been previously considered for other scenarii. In \cite{MV11}, the authors show that the secrecy capacity of classical wiretap channels can be achieved using polar codes. This work was extended to quantum wiretap channels with a classical eavesdropper in \cite{WG12}. In \cite{RDR12}, polar codes are used to transmit quantum information and an efficient decoder is provided for both Pauli channels and erasure channels. In \cite{GW12}, it is shown that the Holevo capacity of lossy optical channels can be achieved with polar codes but an implementation of a quantum successive cancellation decoder is far beyond what can be experimentally realized today with quantum states. 

The QKD and wiretap channel scenarii are nevertheless different: in QKD, Alice and Bob's correlations are directly used to compute an upper bound on Eve's information without making any assumption on the channel between Alice and Eve, whereas in the wiretap channel scenario, the channel between Alice and Eve is assumed to be characterized.

Polar codes exhibit some specificities that make them suitable for QKD error correction. First, they are easily employed in a rateless setup where the noise of the channel can change over time. Secondly, they enable non-interactive error correction, similarly to LDPC codes, and contrary to two-way protocols like Cascade. In this section, we evaluate the block sizes needed to obtain the  efficiencies required for QKD. This impacts the decoding throughputs that can be obtained in practical implementations.

In polar codes, individual copies of symmetric Binary Discrete Memoryless Channels (BDMC) are combined recursively in order to form a new set of channels composed of more and more differentiated channels, such that in the asymptotic limit channels are either error-free or completely noisy, with a fraction of error-free channels equal to the code capacity. This phenomenon is called channel polarization: each channel becomes either noiseless or noisy as the block length goes to infinity. In the asymptotic limit, the capacity of the BDMC can be achieved by sending the information bits through the noiseless channels, while in practice, only a fraction of this capacity is achieved using the  bits with almost zero error probability for finite block lengths. The convergence speed of channels into noiseless or noisy channels is called polarization speed.

\begin{figure}
\centering
 \includegraphics[width=120mm]{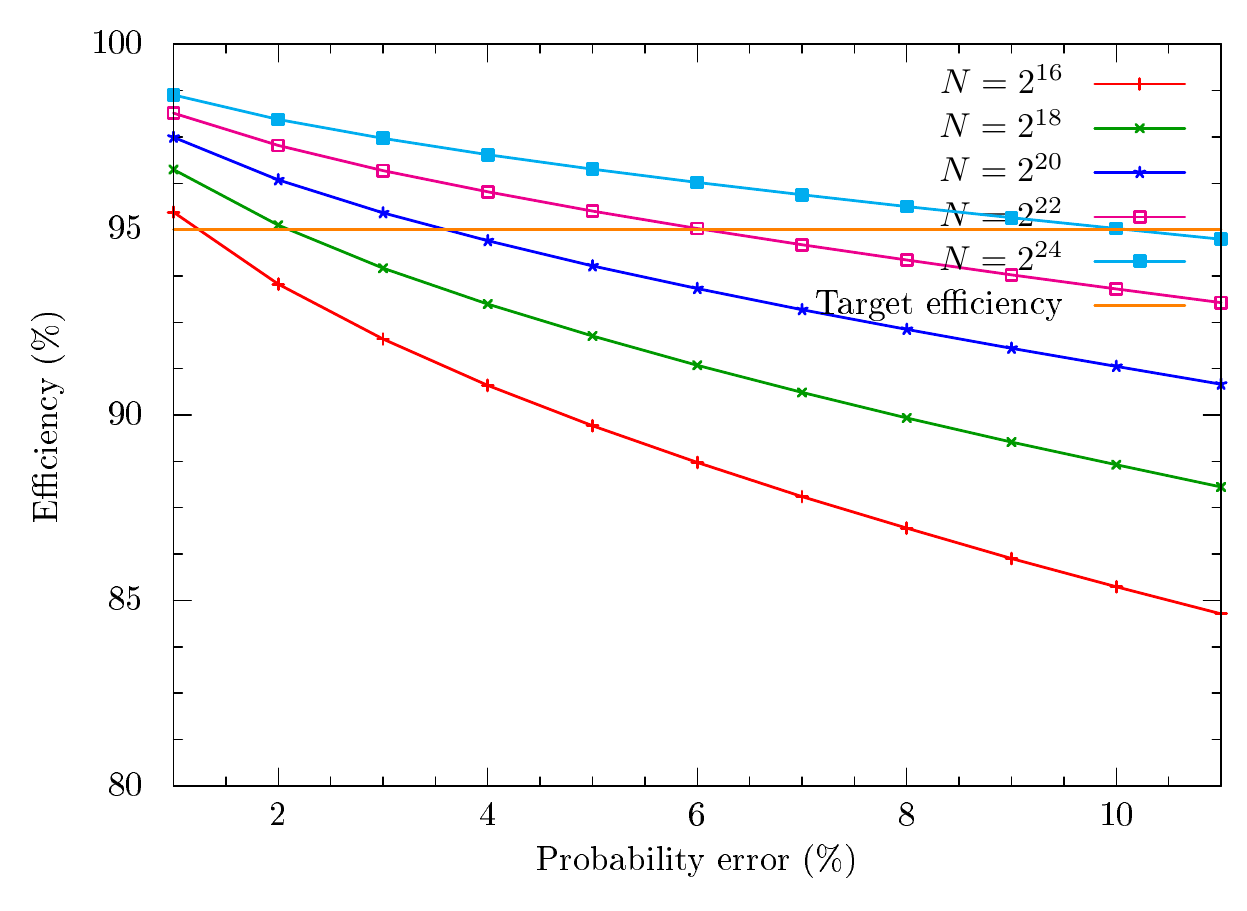}
  \fcaption{Polar codes efficiency for the BSC for probability errors from 1\% to 11\% with a 1\% step. The method described in \cite{MT09} is used to compute the capacities of each channel for a given noise level and the frozen bits are chosen in order to upper bound the FER by 0.1 according to this method. From the bottom to the top we used the following block sizes: $2^{16}$, $2^{18}$, $2^{20}$, $2^{22}$, $2^{24}$. We can see that the efficiency is higher than the target efficiency of 95\% over almost the entire range for block sizes equal to $2^{24}$.}
 \label{figure:polar_bsc}
\end{figure}

\begin{figure}
\centering
 \includegraphics[width=120mm]{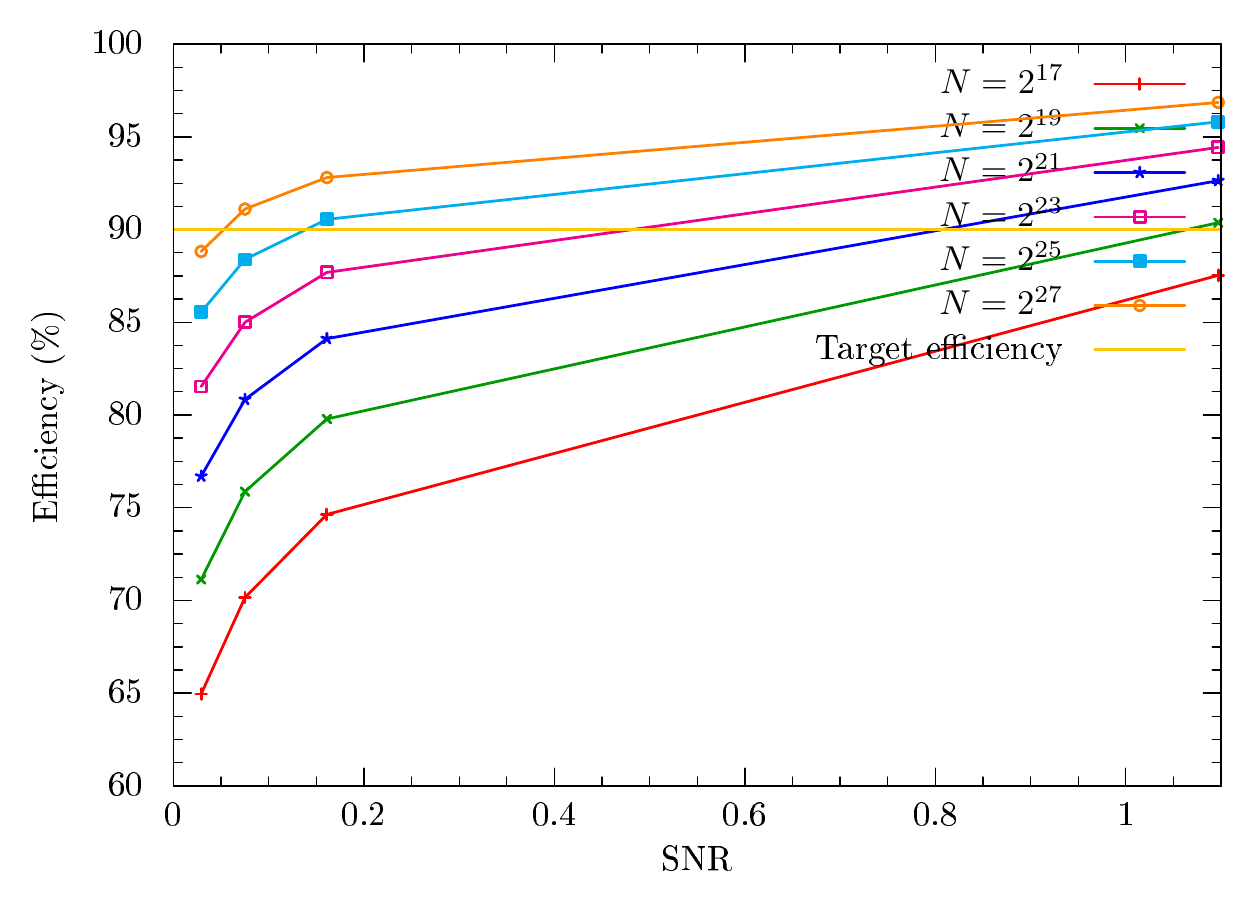}
  \fcaption{Polar codes efficiency for the AWGNC for the SNRs 1.097, 0.161, 0.075, 0.029 from \cite{JKL11}. The method described in \cite{MT09} is used to compute the capacities of each channel for a given noise level and the frozen bits are chosen in order to bound the FER by 0.1 according to this method. From the bottom to the top we used the following block sizes: $2^{17}$, $2^{19}$, $2^{21}$, $2^{23}$, $2^{25}$, $2^{27}$. We can see that the efficiency is higher than the target efficiency of 90\% over almost the entire range for block sizes equal to $2^{27}$.}
 \label{figure:polar_awgnc}
\end{figure}

We used the polar codes construction method described in \cite{MT09} to compute the decoding error probabilities on symmetric binary memoryless channels for the BSC and the BIAWGNC. For a given noise level on a given channel, Density Evolution allows us to compute the capacities of the different bits of the code. Some of the bits corresponding to channels with lowest capacities are simply revealed and are called the frozen bits of the code. As explained in \cite{MT09}, this selection rule for frozen bits also gives us an upper bound on the decoding error probability of a block (also called the Frame Error Rate or FER). Since in QKD it is not crucial to lose some blocks (they will just be thrown away at the verification step), we select sets of frozen bits that give an upper bound of $0.1$ on the FER. It appears that the polarization speed is highly dependent on the channel for polar codes \cite{KOR10}. Figure~\ref{figure:polar_bsc} gives the polarization speed we obtained for the BSC. It shows that polar codes have an efficiency above $95\%$ over almost the entire probability error range $[0;0.11]$, which is the range of interest in DVQKD, for block lengths starting from $2^{24}$. Even smaller block lengths can be used if one does not need to cover the entire probability error range. The situation is definitely worse in Figure~\ref{figure:polar_awgnc} for CVQKD. We studied the polarization speed for the SNRs described in \cite{JKL11} because high efficiency multi-edge LDPC codes have been designed to deal with such noise levels \cite{JKL11, RIC02}. The results show that only a $90\%$ efficiency can be achieved with polar codes for blocks of size $2^{27}$ whereas efficiencies of about $95\%$ are achieved in \cite{JKL11} with LDPC codes. However long distance CVQKD is still possible using polar codes. Furthermore, there is still some hope to improve the polarization speed for polar codes for the BIAWGNC, for example by changing the recursive method used to combine channels, as proposed in \cite{MT10}.

\section{Decoding speed: numerical results}
\label{sec:exp_results}
\begin{table}[t]
\tcaption{Polar codes decoding speeds on the BSC and the BIAWGNC. The efficiencies correspond to a block error rate of 0.1 when selecting the frozen bits according to the method described in \cite{MT09}. These figures were obtained with one core of an Intel Core i5 670 3.47GHz processor.}
\centering
\begin{tabular}{|c|c|c|c|c|c|}
\hline
Channel & QBER / SNR & Size & $\beta$ & Speed (Mb/s) & FER\\
\hline
BSC & 2.0\% & $2^{16}$ & 93.5\% & 10.9 & 0.09\\
BSC & 2.0\% & $2^{20}$ & 96.3\% & 9.5 & 0.11\\
BSC & 2.0\% & $2^{24}$ & 98.0\% & 8.3 & 0.08\\
BIAWGNC & 1.097 & $2^{24}$ & 95.2\% & 8.0 & 0.10\\
BIAWGNC & 0.161 & $2^{27}$ & 92.8\% & 7.3 & 0.09\\
\hline
\end{tabular}
\label{tab:polar_speed}
\end{table}

\begin{table}[t]
\tcaption{LDPC codes decoding speeds with LDPC codes described in \cite{ELK09} for the BSC and in \cite{RIC02} for the BIAWGNC. The maximum number of iterations was fixed to 20 for the first code, and respectively to 160 and 100 for the next two codes. These figures were obtained with an AMD Tahiti Graphics Processor.}
\centering
\begin{tabular}{|c|c|c|c|c|c|c|}
\hline
Channel & QBER / SNR & Size & $\beta$ & Speed (Mb/s) & FER\\
\hline
BSC & 2.0\% & $2^{17}$ & 92.9\% & 7.3 & 0.01\\
BIAWGNC & 1.097 & $2^{20}$ & 96.9\% & 6.5 & 0.09\\
BIAWGNC & 0.161 & $2^{20}$ & 93.1\% & 7.1 & 0.04\\
\hline
\end{tabular}
\label{tab:ldpc_speed_cpu}
\end{table}

\begin{table}[t]
\tcaption{LDPC codes decoding speeds with LDPC codes described in \cite{ELK09} for the BSC and in \cite{RIC02} for the BIAWGNC. The maximum number of iterations was fixed to 15 for the first code, and respectively to 100 and 50 for the next two codes. These figures were obtained with one core of an Intel Core i5 670 3.47GHz processor.}
\centering
\begin{tabular}{|c|c|c|c|c|c|c|}
\hline
Channel & QBER / SNR & Size & $\beta$ & Speed (Mb/s) & FER\\
\hline
BSC & 2.0\% & $2^{17}$ & 93.1\% & 0.82 & 0.03\\
BIAWGNC & 1.097 & $2^{20}$ & 96.9\% & 0.09 & 0.03\\
BIAWGNC & 0.161 & $2^{20}$ & 93.1\% & 0.12 & 0.04\\
\hline
\end{tabular}
\label{tab:ldpc_speed_gpu}
\end{table}

An interesting feature of polar codes is their regular recursive structure. This allows us to implement a recursive, successive-cancellation decoder that achieves a speed of about $10$Mb/s on modern CPUs (Intel Core i5 670 3.47 GHz in the simulations). The main optimization in this decoder is to use fixed-point arithmetic and a table-lookup implementation of the function $\varphi(x)=\operatorname{log}(\operatorname{tanh}(x/2))$ used to update log-likelihood ratios (LLRs). Other techniques have been proposed for efficient polar codes decoding and could improve the decoding speeds given in Table~\ref{tab:polar_speed}: in \cite{ZYP12}, the authors propose look-ahead techniques that allow to reduce the decoding latency of successive cancellation by 50\% while in \cite{TV12, LST12, CNL12}, some variants of list decoding for polar codes are introduced.

The polar decoding performance has to be compared with the speed of a LDPC decoder based on BP. The speed of such a decoder dramatically lowers when approaching the capacity of the code used because BP requires more iterations to converge. Thus LDPC decoding speed is limited to about $800$kb/s using one core of a modern CPU. The LDPC CPU decoder uses fixed-point arithmetic and the same implementation of $\varphi$ as in the polar code case. It is a shuffle decoder with an early termination strategy where bits are considered to be known (and their LLR ceases to be updated) when the absolute value of their LLR passes a threshold; when no bit is updated for a sufficient number of iterations, decoding is considered to be over and is stopped. Because the regime explored is close to the Shannon limit, simplified BP algorithms such as min-sum or its variants cannot be used. Finally, the maximum number of iterations is controlled to adjust the FER to the target value 0.1. This control is imprecise however, since small variations of the maximum allowed number of iterations result in large FER changes. The maximum number of iterations used for LDPC codes are given in Table~\ref{tab:ldpc_speed_cpu} and Table~\ref{tab:ldpc_speed_gpu} legends.

GPUs provide a huge amount of parallelism that allows us to achieve speeds of $10$Mb/s (figures are given for an AMD Tahiti Graphics Processor). The GPU LDPC decoder is different from the CPU implementation: it is a floating-point, flood decoder running in a fixed number of iterations and using both 'external' parallelism (several vectors are decoded concurrently) and 'internal' paralelism (for a single BP execution corresponding to one message being decoded, several messages are propagated concurrently). This was experimentally found to be optimal on GPU architectures because they have much more floating point computational power than CPUs, but are slowed down by complex control logic. No competitive GPU decoder for polar codes was implemented, as successive cancellation is inherently sequential, and therefore only external parallelism can be used.

Table~\ref{tab:polar_speed} gives the decoding speeds obtained with polar codes for the BSC and the BIAWGNC for characteristic noise levels in DVQKD and CVQKD. Table~\ref{tab:ldpc_speed_gpu} and Table~\ref{tab:ldpc_speed_cpu} give the corresponding speeds with LDPC codes respectively with a GPU and a CPU. 

The best reported QKD key rate is about $1$Mb/s \cite{SAS11, DYDSS10} (which is several order of magnitudes below state-of-the-art optical communication links that range from $1$Gb/s to $100$Gb/s). This means that even using large blocks as in Table~\ref{tab:polar_speed}, polar codes decoding throughput is enough for state-of-the-art QKD implementations.

\section{Conclusion}
\noindent
We showed that polar codes can be used to perform the error correction step for both DVQKD and CVQKD. They achieve good efficiencies for the BSC and BIAWGNC for level of noises compatible with QKD. However, since the polarization speed of polar codes is worse for the BIAWGNC than for the BSC, they require higher block sizes and are less practical for CVQKD than for DVQKD. 

As regards the decoding step, which is often a bottleneck in recent QKD implementations, we showed that polar codes feature high-speed recursive decoding and achieve CPU decoding speeds similar to LDPC GPU decoding speeds. This is to our knowledge the first practical application of polar codes.

\nonumsection{Acknowledgements}
\noindent
This research was supported by the French National Research Agency, through the FREQUENCY (ANR-09-BLAN-0410) and HIPERCOM projects, and by the European Union, through the project Q-CERT (FP7-PEOPLE-2009-IAPP).
P. Jouguet acknowledges support from the ANRT (Agence Nationale de la Recherche et de la Technologie). The authors thank J.C. Belfiore for fruitful discussions.

\nonumsection{References}

\end{document}